\documentstyle[12pt,epsf]{article}

\begin{document}

\baselineskip 24pt

\newcommand{\sheptitle}
{Flavour Democracy in Strong Unification} 

\newcommand{\shepauthor}
{S. A. Abel and S. F. King
\footnote{On leave of absence from 
Department of Physics and Astronomy,
University of Southampton, Southampton, SO17 1BJ, U.K.}}

\newcommand{\shepaddress}
{Theory Division, CERN, CH-1211 Geneva 23, Switzerland.}

\newcommand{\shepabstract}
{We show that the fermion mass spectrum may naturally be understood in terms of
flavour democratic fixed points
in supersymmetric theories which have a large domain of attraction
in the presence of ``strong unification''.
Our approach provides an alternative to the approximate Yukawa texture zeroes
of the Froggatt-Nielsen mechanism.
We discuss a particular model based on a broken 
gauged $SU(3)_L\times SU(3)_R$ family symmetry which illustrates our approach.}

\begin{titlepage}
\begin{flushright}
CERN-TH/98-144\\
hep-ph/9804446\\
\end{flushright}
\vspace{0.5in}
\begin{center}
{\large{\bf \sheptitle}}
\bigskip \\ \shepauthor \\ \mbox{} \\ {\it \shepaddress} \\ 
\vspace{0.5in}
{\bf Abstract} \bigskip \end{center} \setcounter{page}{0}
\shepabstract
\vspace{0.5in}
\begin{flushleft}
CERN-TH/98-144\\
\today
\end{flushleft}
\end{titlepage}

\def\sspace{\baselineskip = .16in}
\def\dspace{\baselineskip = .30in}
\def\beq{\begin{equation}}
\def\eeq{\end{equation}}
\def\bea{\begin{eqnarray}}
\def\eea{\end{eqnarray}}
\def\bq{\begin{quote}}
\def\eq{\end{quote}}
\def\ra{\rightarrow}
\def\lra{\leftrightarrow}
\def\ups{\upsilon}
\def\bq{\begin{quote}}
\def\eq{\end{quote}}
\def\ra{\rightarrow}
\def\un{\underline}
\def\ov{\overline}

\newcommand{\plb}[3]{{{\it Phys.~Lett.}~{\bf B#1} (#3) #2}}
\newcommand{\npb}[3]{{{\it Nucl.~Phys.}~{\bf B#1} (#3) #2}}
\newcommand{\phrd}[3]{{{\it Phys.~Rev.}~{\bf D#1} (#3) #2}}
\newcommand{\ptp}[3]{{{\it Prog.~Theor.~Phys.}~{\bf #1} (#3) #2}}
\newcommand{\leqsim}{\,\raisebox{-0.6ex}{$\buildrel < \over \sim$}\,}
\newcommand{\geqsim}{\,\raisebox{-0.6ex}{$\buildrel > \over \sim$}\,}
\newcommand{\be}{\begin{equation}}
\newcommand{\ee}{\end{equation}}
\newcommand{\ba}{\begin{eqnarray}}
\newcommand{\ea}{\end{eqnarray}}
\newcommand{\nn}{\nonumber}
\newcommand{\ie}{\mbox{{\em i.e.~}}}
\newcommand{\mpl}{\mbox{$M_{pl}$}}
\def\gev{\,{\rm GeV}}

\section{Introduction}

Attempts to understand the pattern of fermion masses
and mixing angles in supersymmetric theories are usually
based on the idea of Yukawa texture zeroes, or approximate
texture zeroes which result from high powers of a small expansion
parameter as in the Froggatt-Nielsen approach~\cite{frog}.
In this paper we discuss an alternative to the texture approach
based on the idea of flavour democracy~\cite{democra}.
If flavour democracy is implemented in conventional supersymmetric 
grand unified theories (SUSY GUTs) then the perturbations 
from democracy required at the GUT scale
in order to account for light masses and mixing angles must be
tuned to be very small. This is because the renormalisation group
equations (RGEs) of the minimal supersymmetric standard model (MSSM)
tend to cause any small deviations from democracy to be magnified when
the Yukawa couplings are run down to low energy.
In this paper we point out that in a particular class of supersymmetric
unified model, namely those with ``strong unification'' and a particular
Higgs structure, the situation is reversed and arbitrary Yukawa
matrices at high energy get driven to flavour democratic fixed points
at low energies.
In such theories the fermion mass spectrum that we
observe in low energy experiments containing all the familiar 
(but bizarre) hierarchies 
may result from rather arbitrary Yukawa matrices at high energies.

Flavour democratic fixed points are a 
natural consequence of ``strong unification''  in which
there is extra matter in complete `$SU(5)$' representations at an
intermediate mass scale $M_{I}$ below the unification scale \cite{strong1}.
The extra matter consists of $n_{5}$ copies of $(5+\bar{5})$ plus $n_{10}$
copies of $(10+\bar{10})$ representations which serve to increase the beta
functions above the scale $M_{I}$, resulting in an increased value of the
unified gauge coupling $\alpha _{GUT}$ \cite{strong1}. On the other hand the
unification scale $M_{GUT}\sim 2\times 10^{16}$ GeV is virtually unchanged
from its MSSM value due to an accurate cancellation between the two-loop and
threshold effects \cite{strong1}. The presence of such additional matter is
typical of a certain class of string model in which gauge symmetries are
broken by Wilson lines \cite{strong2}. Moreover such extra matter is welcome
since it may serve to increase the unified coupling to a value which is
large enough to solve the ``dilaton runaway problem'' \cite{dilaton}, 
provided that 
the string scale $M_{string}$ is reduced down to $M_{GUT}$ as suggested 
in Ref.~\cite{W}.

In such theories the values of the gauge couplings near the unification
scale may be raised into the strong coupling region \cite{strong3},
effectively placing the question of the unification of the gauge couplings
outside perturbation theory. At first sight this would seem to imply that
all the predictive power of unification is lost. However, as shown in \cite
{strong3}, low energy predictivity is maintained since the steeply falling
gauge couplings are quickly driven to precise fixed point ratios: 
\begin{equation}
\frac{\alpha _{1}}{\alpha _{3}}\rightarrow r_{1}\equiv \frac{b_{3}}{b_{1}},\ \
\frac{\alpha _{2}}{\alpha _{3}}\rightarrow r_{2}\equiv \frac{b_{3}}{b_{2}},  
\label{r}
\end{equation}
where the beta functions are 
\begin{equation}
b_{a}=\left( 
\begin{array}{c}
33/5+n \\ 
1+n \\ 
-3+n
\end{array}
\right)  \label{b}
\end{equation}
where $n=(n_{5}+3n_{10})/2$. Thus one may take the ratios $r_{1},r_{2}$ as a
boundary condition at the scale $M_{I}$, and use them to determine the
low energy measured couplings. In this approach the scale $M_{I}$ is
regarded as an input parameter which may, for a given value of $n$, be fixed
by two of the gauge couplings (say $\alpha _{1}$ and $\alpha _{2}$). The
third gauge coupling may be predicted at low energies as in the standard
unification picture, and indeed leads to values of $\alpha _{3}(M_{Z})$ in
good agreement with experiment \cite{strong3}. This prediction, which
follows without a conventional scale $M_{GUT}$, originates from the precise
boundary conditions in Eq.(\ref{r}) at $M_{I}$. The gauge couplings become
non-perturbative at a scale $M_{NP}\sim 3\times 10^{16}$ GeV, close to the
conventional GUT scale \cite{strong3}. Note that $M_{I}$ is the mass scale
in the superpotential, {\it not} the physical mass of the heavy states which
receive large radiative corrections. Such radiative splitting effects
decouple from the evolution equations for the couplings~\cite{shifman}.
The key to the predictive power of this scheme is the steeply falling gauge
couplings in the region $M_{NP}-M_{I}$, which drives the gauge couplings to
their fixed point values at $M_{I}$ in Eq.(\ref{r}) \cite{strong3}. 

Similar fixed points also apply to the Yukawa couplings whose fixed points
are of the Pendleton-Ross \cite{PR} type rather than the quasi-fixed
point type \cite{H}.\footnote{For a discussion of the relation between these
two types of fixed point see \cite{LR}.}
It was shown that if all the third family Yukawa couplings are assumed to be
in the (large) domain of attraction of the fixed point at high energies then
this leads to precise predictions for third family Yukawa couplings at the
scale $M_{I}$ and hence to precise low energy predictions for third family
masses and the ratio of Higgs vacuum expectation values $\tan \beta $ as a
function of the parameter $n$ \cite{strong3}. In this case $\tan \beta
\approx 46-47$ and the top quark mass exceeds 200 GeV in all cases \cite
{strong3}. However, as pointed out,
these predictions for the third family Yukawa couplings are sensitive to
other Yukawa couplings which are large enough to be within the domain of
attraction of the fixed point and, in a particular theory of fermion 
masses~\cite{GG}, 
the presence of large Yukawa couplings involving the first and
second families will affect the low energy predictions of the third family
spectrum and reduce the top mass prediction to acceptable values \cite
{strong3}.

In one implementation of the model of ref.\cite{GG} there is a separate Higgs 
doublet for each entry of the Yukawa matrix, leading to
9 Higgs doublets $H_{U_{ij}}$ coupling to the up sector and 9 Higgs
doublets $H_{D_{ij}}$ coupling to the down and lepton sector.
The Yukawa matrices just above $M_I$ are of the flavour democratic
form as a consequence of the fixed point structure of the theory.
However the flavour democracy is destroyed by the manner in which the
two low energy Higgs doublets of the MSSM $H_U$ and $H_D$
are extracted from the 18 Higgs doublets which couple to quarks and leptons.
Essentially the light doublets are identified as
$H_U\sim  H_{U_{33}}$ and $H_D\sim H_{D_{33}}$, with all the other
Higgs acquiring mass of order $M_I$. Thus, from the point of view of the
MSSM, the two Higgs doublets only have large Yukawa couplings
in the 33 entry and, since these Yukawa couplings are roughly equal
at the fixed point,
one must explain the top-bottom mass hierarchy by taking a large
ratio of low energy Higgs vacuum expectation values
$\tan \beta \sim m_t/m_b$.
Other entries of the Yukawa matrix generated
by Higgs mixing effects controlled by 
a gauged $U(1)_{X}$ family symmetry\cite{IR}. The $U(1)_{X}$ has a
Green-Schwarz anomaly \cite{GS} and is assumed to be broken close to the
string scale by the vacuum expectation values (VEVs) of standard model
singlet fields $\theta $ and $\bar{\theta}$ with $U(1)_{X}$ charges 1 and -1
respectively \cite{IR}. In order to achieve a realistic pattern of masses
the simple $X$ charges are assumed for the three families of
quarks and leptons.
The 9 Higgs of each type mix via
Froggatt-Nielsen \cite{frog} diagrams involving insertions 
of the $\theta $ and 
$\bar{\theta}$ fields along the Higgs line, so that at low energies
effective Yukawa matrices emerge. Approximate texure zeroes and 
hierarchies are interpreted as high powers of an expansion parameter
\beq
\epsilon =<\theta >/M_{I}=<\bar{\theta}>/M_{I}
\label{epsilon}
\eeq
where $\epsilon \approx 0.2$.

In the above model \cite{GG} it is clear that the flavour democracy
of the high energy theory at its fixed point is not transmitted to the
MSSM since it is maximally broken by the way in which the light Higgs
doublets are identified. However the embedding of $H_U$ and $H_D$
in the high energy theory is certainly not unique; for example it is
possible to identify $H_U\sim H_{U_{33}}$ but 
$H_D\sim H_X + \gamma H_{D_{33}}$ where $H_X$ is some extra Higgs 
doublet which does not couple to quarks and leptons and
$\gamma \sim m_b/m_t$ allows values of $\tan \beta \sim 1$.
Indeed one can envisage further possibilities for embedding
the MSSM Higgs in the high energy theory. In this paper we are interested
in the possibility that the MSSM Higgs doublets preserve the
flavour democracy of the high energy theory.
In other words we shall suppose 
that the light Higgs doublets of the MSSM are
democratic mixtures of the Higgs doublets in the high energy theory:
\begin{eqnarray}
H_U & = &  \frac{1}{3}\sum_{ij=1,3} H_{U_{ij}} \nonumber \\
H_D & = &  \frac{1}{3}\sum_{ij=1,3} H_{D_{ij}}.
\label{democraticHiggs}
\end{eqnarray}
In this case the flavour democratic Yukawa couplings of the high energy
theory will be preserved in matching the theory onto the MSSM, and
we will have realised our goal of obtaining flavour democracy as
an infrared fixed point.

How can the democratic combinations in Eq.\ref{democraticHiggs} be achieved
in practice? One simple example is to take the $\epsilon \rightarrow 1$
limit of the above theory \cite{GG}. In this limit we would expect the
Higgs mixing to be maximal corresponding to the approximately
democratic combinations in Eq.\ref{democraticHiggs}.
In the usual model the Higgs $H_{U_{ij}}$ and $H_{D_{ij}}$ are all assigned
various $X$ charges consistent with their renormalisable
couplings to the quarks and leptons which are assigned $X$ charges
of $-4,1,0$ for the 1st, 2nd, 3rd families. The  $H_{U_{33}}$ and
$H_{D_{33}}$ are therefore assigned zero $X$ charge. Furthermore the
18 Higgs doublets above are accompanied by 16 conjugate Higgs doublets
$\bar{H}_{U_{ij}}$ and $\bar{H}_{D_{ij}}$ (with $i=j=3$ missing)
which carry opposite quantum numbers and so form vector masses $M_I$ with
16 of the Higgs doublets. Since 
$\bar{H}_{U_{33}}$ and $\bar{H}_{D_{33}}$ are missing
this implies that ${H}_{U_{33}}$ and ${H}_{D_{33}}$ remain light and so
are identified as the two Higgs doublets of the MSSM.
This is just the usual scheme in which a small effective Yukawa coupling
in the $ij$ position of the up-matrix for example
is small because the corresponding Higgs ${H}_{U_{ij}}$
has some $X$ charge which must be stepped down via $n$ insertions
of $\theta$ fields to reach the physical Higgs ${H}_{U_{33}}$
which has zero $X$ charge, causing the effective Yukawa
coupling to be of order $\epsilon^n$. Now if $\epsilon \approx 1$
there is no price to pay for $\theta$ field insertions,
so the effective Yukawa couplings in all the entries would be expected to be
approximately equal to their fixed point values in the
high energy theory. In other words the Higgs would mix democratically
as in Eq.\ref{democraticHiggs}. Of course the precise nature of the
mixing depends also on the full set of Yukawa couplings which 
control the mixing of the singlets with the Higgs, which are rather
complicated \cite{ben}. In general it is rather difficult to get
a ``handle'' on this kind of approach to the democratic Higgs, so
we now introduce an alternative mechanism which
departs from the $X$ symmetry completely. 

\section{A New Model of Democratic Higgs}

The real explanation for the democratic Higgs may lie in 
additional fixed points of the high energy theory~\cite{ben}
leading to the picture described above, or it may have to do with 
the non-perturbative physics existing above the scale $M_{NP}$. 
For the moment however we would like to have a working model 
of flavour democracy. In this section we will therefore present 
a scheme in which a flavour democratic Higgs is generated 
in a perturbative supergravity framework, bearing in mind that 
non-perturbative effects could in principle have important effects
which are beyond our analysis.

In order to present a perturbative model, we are obliged 
to depart significantly from the model of Ref.\cite{GG}.
In the example we consider in this section there 
is an additional $SU(3)_L\times SU(3)_R$ family 
symmetry at the Planck scale, under which all the fields 
transform;
\ba
Q &=& (\overline{3},1)\nonumber\\
L &=& (\overline{3},1)\nonumber\\
U^c &=& (1,3)\nonumber\\
D^c &=& (1,3)\nonumber\\
E^c &=& (1,3)\nonumber\\
\nu^c &=& (1,3)\nonumber\\
H_U &=& (3,\overline{3})\nonumber\\
H_D &=& (3,\overline{3})\nonumber\\
D_U &=& 8\times(1,1)\nonumber\\
D_D &=& 8\times(1,1).
\ea
We have added the 16 extra `down' superfields so that 
the content in addition to the MSSM falls into $5+\overline{5}$ 
multiplets as required by the renormalisation group running 
from the intermediate scale although the symmetry we have chosen means 
that there can no longer be any underlying $SU(5)$ symmetry\footnote{ 
The requirement is that $\alpha_s$ should be correct 
when we run the gauge couplings down from the boundary conditions 
at the intermediate scale which are dictated by the fixed points. If we 
are prepared to drop the `unification' normalisation of the 
$U(1)_Y$ charges, $k_1=5/3$, then other solutions are possible -- 
although there are no solutions which require no extra `down' states.}.
We have also added a right handed neutrino to cancel any 
potential anomalies. 

In order to obtain the democratic Higgs we need to add 
extra fields which are singlets under the Standard Model
gauge group but which transform under the $SU(3)_L\times SU(3)_R$ 
family symmetry.
The family symmetry is broken by Planck scale VEVs of four 
fields which transform in the adjoint representation of each group 
factor (flagged by the subscripts);
\ba
\Omega_L &=& (8,1)\nonumber\\
\Omega'_L &=& (8,1)\nonumber\\
\Omega_R &=& (1,8)\nonumber\\
\Omega'_R &=& (1,8)
\ea
where 
\be
\langle \Omega_L \rangle \mbox{ ,}
\langle \Omega_R \rangle \mbox{ ,}
\langle \Omega'_L \rangle \mbox{ ,}
\langle \Omega'_R \rangle = {\cal{O}}(1)
\ee
in natural units. (The VEVs 
must also commute, which is a mild assumption if there are non-trivial 
interactions in $D$-terms for example.)
In addition we need gauge singlet fields to generate the 
intermediate scale, $M_I$, and select the democratic Higgs to be 
the low energy (below $M_I$) Higgs. These are 
\ba
\Theta_L &=& (3,1)\nonumber\\
\Theta'_L &=& (\overline{3},1)\nonumber\\
\Theta_R &=& (1,3)\nonumber\\
\Theta'_R &=& (1,\overline{3})\nonumber\\
\Lambda_L &=& (6,1)\nonumber\\
\Lambda'_L &=& (\overline{6},1)\nonumber\\
\Lambda_R &=& (1,6)\nonumber\\
\Lambda'_R &=& (1,\overline{6})\nn\\
S_R, S_L, S'_R, S'_L &=& (1,1)
\ea
Finally, in order to enforce the correct form of 
couplings we invoke an extra discrete $Z_N$ symmetry
under which the non-zero charges are
\ba 
Z_{\Omega_L} &=& 1 \nn\\ 
Z_{S_L}=Z_{L}=Z_{Q} &=& -1,
\ea
a $Z'_{N}$ symmetry with non-zero charges
\ba
Z'_{\Omega_R} &=& 1 \nn\\
Z'_{S_R}=Z'_\nu=Z'_{E^c}=Z'_{U^c}=Z'_{D^c} &=& -1,
\ea
a $Z''_N$ symmetry under which the non-zero charges are 
\ba
Z''_{\Omega'_R} = Z''_{\Omega'_L} &=& 1 \nn\\
Z''_{S'_R}=Z''_{S'_L} &=& -1,
\ea
and a $Z'''_N$ symmetry under which the non-zero charges are 
\ba
Z'''_{\Theta_L} = Z'''_{\Theta'_R} = Z'''_{\Lambda_L} = 
Z'''_{\Lambda'_R} &=& 1 \nonumber\\
Z'''_{\Theta'_L} = Z'''_{\Theta_R} = Z'''_{\Lambda'_L} = 
Z'''_{\Lambda_R} &=& -1.
\ea
With this set of charges the superpotential is of the
correct form to give us the low energy democratic Higgs structure we 
require. The most general superpotential allowed by the above 
symmetries is  
\be
\label{spot1}
W=W_{yuk}+W_\mu+W_D+W_S
\ee
where
\ba
W_{yuk}&=& %
\lambda_u  Q \Omega_L H_U \Omega_R U^c +
\lambda_{d}  Q \Omega_L H_D  \Omega_R D^c \nn\\ &&\hspace{2cm}
+ 
\lambda_\nu  L \Omega_L H_U \Omega_R \nu^c +
\lambda_{e}  L \Omega_L H_D  \Omega_R E^c +\ldots,
\label{spot2}\\
W_\mu &=&
\lambda_{\mu_R} \varepsilon \Theta_L \Lambda_R H_U H_D+
\lambda_{\mu_L} \varepsilon \Theta'_R \Lambda'_L H_U H_D+\ldots
\label{spot3}
\nn\\ 
W_D &=& \lambda_{D_L}\Theta_L \Theta'_L D_D D_U +  
\lambda_{D_R}\Theta_R \Theta'_R D_D D_U +\ldots
\ea
and 
\ba
W_S &=& S_L \Theta'_L \Omega_L \Theta_L
+S_R \Theta'_R \Omega_R \Theta_R
+S'_L \Theta'_L \Omega'_L \Theta_L
+S'_R \Theta'_R \Omega'_R \Theta_R
\nn\\ &&\hspace{1cm}
+\lambda_{\Theta_L} (\Theta'_L\Theta_L)^2
+\lambda_{\Theta_R} (\Theta'_R\Theta_R)^2
+\lambda_{\Lambda_L} (\Lambda'_L\Lambda_L)^2
+\lambda_{\Lambda_R} (\Lambda'_R\Lambda_R)^2+\ldots.
\label{spot4}
\ea
In the above the ellipses indicate higher order terms which 
are suppressed by at least a factor of $\Omega_L^{N} $ or $\Omega_R^{N}$
or $\Theta_{L,R},\Theta_{L,R}', \Lambda_{L,R},\Lambda_{L,R}' $. (All of 
the latter get small VEVs as we shall shortly see.
We are also assuming that there are no mass terms for these fields.) 
Since we assume $\langle \Omega \rangle < 1 $ (in natural units) 
it is safe to neglect them provided that $N$ is a large number. 
The $\varepsilon $'s are Levi-Cevita symbols
for the family symmetry with $SU(3)$ indices being suppressed; 
hopefully the contractions are self evident. 
The $\lambda$'s are single couplings.

We now go through the superpotential term by term to 
describe the role each piece plays. The first term, $W_{yuk}$ 
leads to a Yukawa coupling structure ansatz which is quite 
restrictive although remarkably successful; in particular we 
will see later that the mass matrices
have two massless eigenvalues at the $M_{NP}$ 
scale, leading to the required CKM and mass structure
by the time the model is renormalised down to $M_I$. (With this 
ansatz there are only 5 free parameters in the Yukawa couplings;
initially there are 7, 
($\lambda_{u}$, $\lambda_{d}$, $\lambda_{e}$, 
$\langle \Omega_L \rangle_{11}$, $\langle \Omega_L \rangle_{22}$, 
$\langle \Omega_R \rangle_{11}$ and $\langle \Omega_R \rangle_{22}$),
but $\lambda_u$ and $\Omega_{L11}$ may be absorbed into the definition 
of the other parameters.) 
Indeed the Yukawa couplings at 
$M_{NP}$ can be identified as
\ba
\label{cups}
h_{ij}&=& \lambda_{u}
\langle \Omega_L\rangle_{ii} \langle \Omega_R \rangle_{jj} 
\nn\\
k_{ij}&=& \lambda_{d}
\langle \Omega_L\rangle_{ii} \langle \Omega_R \rangle_{jj} 
\nn\\
l_{ij}&=& \lambda_{e}
\langle \Omega_L\rangle_{ii} \langle \Omega_R \rangle_{jj} 
\ea
once we have rotated to a basis in which the adjoint VEVs are 
diagonal. 

The VEVs of the $\Theta $ fields are enforced by the $W_S$ term 
to be democratic as long as $\langle S_L\rangle = \langle S_R\rangle = 
\langle S'_L\rangle = \langle S'_R\rangle = 0$;
the $F$-flatness condition is 
\ba
F_{S_L}&=& \mbox{Tr} \Theta'_L \Omega_L\Theta_L =0  \nn\\
F_{S_R}&=& \mbox{Tr} \Theta'_R \Omega_R\Theta_R =0  \nn\\
F_{S'_L}&=& \mbox{Tr} \Theta'_L \Omega'_L\Theta_L =0  \nn\\
F_{S'_R}&=& \mbox{Tr} \Theta'_R \Omega'_R\Theta_R =0
\ea 
which imposes 
\ba
\langle \Theta_{L_1} \Theta'_{L_1} \rangle &=&
\langle \Theta_{L_2} \Theta'_{L_2} \rangle =
\langle \Theta_{L_3} \Theta'_{L_3} \rangle \nn\\ 
\langle \Theta_{R_1} \Theta'_{R_1} \rangle &=&
\langle \Theta_{R_2} \Theta'_{R_2} \rangle =
\langle \Theta_{R_3} \Theta'_{R_3} \rangle 
\ea
as a result of the tracelessness of the adjoint VEVs. 
In supergravity, the scalar potential is of the form 
\be
V=- e^{K} 
\left( 
3 |W|^2 - 
(K_i W +W_i)
K^{i\overline{j}}
(K_{\overline{j}}\overline{W}+ \overline{W}_{\overline{j}}
)\right)
\ee
where $K$ is the K\"ahler potential, $i,j$ label generic fields,
subscripts imply differentiation, and 
$K^{i\overline{j}}=K^{-1}_{\overline{j}i}$. We look for non-trivial 
solutions which are $F$-flat and therefore represent minima 
of the potential. 
They are given by the solutions to
\be
\label{gravy}
(K_i \langle W\rangle  +W_i)=0 
\ee
where $\langle W \rangle \sim m_W $ is fixed by the requirement 
that supersymmetry breaking in the visible sector be of order 
$m_W$. By defining the K\"ahler potential to be minimal
(these can be considered to be the first terms in an expansion) 
\be 
K=
 \Theta_L \overline{\Theta}_L+
 \Theta'_L \overline{\Theta}'_L+
 \Theta_R \overline{\Theta}_R+
 \Theta'_R \overline{\Theta}'_R+
 \Lambda_L \overline{\Lambda}_L+
 \Lambda'_L \overline{\Lambda}'_L+
 \Lambda_R \overline{\Lambda}_R+
 \Lambda'_R \overline{\Lambda}'_R+ \ldots
\ee
we see that Eq.(\ref{gravy}) implies that 
\ba
\langle \Theta_L \rangle=\langle \Theta'_L \rangle 
&=& \theta_L (1,1,1)\nn\\
\langle \Theta_R \rangle =\langle \Theta'_R \rangle 
&=& \theta_R (1,1,1)
\ea
where 
\be
\langle\theta_{L}\rangle,~
\langle\theta_{R}\rangle \sim (m_W/\lambda)^{1/2}
\ee
in natural units, where $\lambda $ is one of 
$\lambda_{\Theta_L},~\lambda_{\Theta_R} $,
and also that the VEVs of the $\Lambda $ fields are of the same order.
(In fact this situation holds even for the most general K\"ahler 
potential.)
The $W_\mu $ term now generates mass terms for Higgs fields of order 
\be
M_I \sim 
\lambda_{\mu_R}m_W/\lambda_{\Theta_L}+
\lambda_{\mu_L}m_W/\lambda_{\Theta_R}
\ee
for all except for the democratic Higgs which remains light; this can be 
seen from the fact that for example 
\be 
\varepsilon \langle\Theta_L\rangle =\theta_L\left(
\begin{array}{rrr}
0 & 1 & -1\nn\\
-1 & 0 & 1\nn\\
1 & -1 & 0
\end{array}
\right).
\ee
This matrix has one eigenvector with zero eigenvalue -- the 
democratic one, $(1,1,1)$. The first term of $W_\mu $ gives masses 
of order $M_I$ to all components of $H_U$ and $H_D$ which 
are not democratic in left indices, and the second term to all 
components which are not democratic in right indices. Thus 
the only component of the Higgs fields which does {\em not} 
receive a mass from the 
first and second terms is that which is democratic in both 
left and right indices -- \ie the democratic Higgs 
(as may easily be checked by expanding out 
and finding the zero eigenvector of the full $9\times 9$ Higgs mass matrix). 
This, low energy Higgs can recieve mass from the 
higher order contributions which should be of order $m_W$ for 
phenomenology. These terms can be at most  
of order $\Omega_{L,R}^N$ so that we require 
$\langle \Omega_{L,R}\rangle^N
\sim m_W/M_I \sim \lambda_{\Theta_L,\Theta_R}$ in order to generate 
the conventional $\mu$-term of the MSSM.
They also disturb the democratic Higgs thereby introducing a 
mixing into the CKM matrix of order $m_W/M_I\sim \lambda_{\Theta_L,\Theta_R}$.
Thus we require $\lambda_{\Theta_L,\Theta_R}\leqsim 10^{-4}$ and hence 
$M_I \geqsim 10^{6}\gev$ to avoid generating significant mixing this way. 

Thus with the set of multiplets and charges
defined above, a democratic Higgs results in perturbative 
supergravity. 
This model has of course no other justification, but its existence, 
and the Standard Model like structure 
which (as we shall see) results, makes Higgs democracy in strong 
unification an avenue worth exploring.
This and the model outlined in the previous section
should therefore be thought of as an
existence proof of the possibility of obtaining a democratic Higgs.
In the following section we describe the renormalisation of the 
strongly unified model, and then we go on to 
show how the Standard Model like structure emerges during the 
running of the high energy theory above the scale $M_I$.

\section{Renormalisation Group Equations}
In the region below the scale $M_{NP}$,
the scale at which the gauge couplings become large,
but above $M_{I}$, the renormalisable superpotential
contains Yukawa couplings involving the 18 Higgs doublets $H_{U_{ij}}$,$%
H_{D_{ij}}$ as well as `$\mu$-terms' which couple the Higgs to each other
and which can generically be of order $M_I$. 
These are 
\ba
W &=& %
\sum_{i,j=1}^{3}\mbox{\huge{(}}h_{ij}Q_{i}U_{j}^{c}H_{U_{ij}}
+k_{ij}Q_{i}D_{j}^{c}H_{D_{ij}}+l_{ij}L_{i}E_{j}^{c}H_{D_{ij}}
\nn\\
& & \mbox{\hspace{7cm}}+ \mu_{ijkl} H_{U_{ij}} H_{D_{kl}}
\mbox{\huge{)}}.
\label{W}
\ea
Such a structure leads to flavour symmetric or democratic Yukawa
fixed points for the Yukawa couplings $h_{ij}$, $k_{ij}$, $l_{ij}$, 
\cite{GG}.\footnote{The soft mass RGEs corresponding to this theory
have recently been studied in \cite{KR}.}

The renormalisation group equations (RGEs) for the gauge couplings are:
\begin{equation}
\frac{d\tilde{\alpha}_{a}}{dt}=-b_{a}\tilde{\alpha}_{a}^{2},
\end{equation}
where we have defined 
$\tilde{\alpha}_{a}\equiv \frac{g_{a}^{2}}{16\pi ^{2}}$, 
$t\equiv \ln (M_{NP}^2/\mu ^{2})$ 
with $\mu $ being the $\bar{MS}$ scale and $b_{a}$
the beta functions given in Eq.(\ref{b}). 
The RGEs for the Yukawa couplings may be factorised into a Yukawa coupling
multiplied by a sum of wavefunction anomalous dimensions for the three legs: 
\begin{eqnarray}
\frac{dY^{h_{ij}}}{dt} &=&Y^{h_{ij}}(N_{Q_{i}}+N_{U_{j}^{c}}+N_{H_{U_{ij}}})
\nonumber \\
\frac{dY^{k_{ij}}}{dt} &=&Y^{k_{ij}}(N_{Q_{i}}+N_{D_{j}^{c}}+N_{H_{D_{ij}}})
\nonumber \\
\frac{dY^{l_{ij}}}{dt} &=&Y^{l_{ij}}(N_{L_{i}}+N_{E_{j}^{c}}+N_{H_{D_{ij}}})
\label{YukRGEs}
\end{eqnarray}
where we have defined $Y^{h_{ij}}\equiv \frac{h_{ij}^{2}}{16\pi ^{2}}$, $%
Y^{k_{ij}}\equiv \frac{k_{ij}^{2}}{16\pi ^{2}}$, $Y^{l_{ij}}\equiv \frac{%
l_{ij}^{2}}{16\pi ^{2}}$. If we assume that the gauge couplings are rapidly
driven to their fixed point ratios then the wavefunction anomalous
dimensions, $N_{i\ }$ may be expressed in terms of the single gauge
coupling $\tilde{\alpha}_{3}$ as: 
\begin{eqnarray}
N_{Q_{i}} &=&(\frac{8}{3}+\frac{3}{2}r_{2}+\frac{1}{30}r_{1})\tilde{\alpha}%
_{3}-\sum_{j=1}^{3}(Y^{h_{ij}}+Y^{k_{ij}})  \nonumber \\
N_{U_{i}^{c}} &=&(\frac{8}{3}+\frac{8}{15}r_{1})\tilde{\alpha}%
_{3}-2\sum_{j=1}^{3}Y^{h_{ji}}  \nonumber \\
N_{D_{i}^{c}} &=&(\frac{8}{3}+\frac{2}{15}r_{1})\tilde{\alpha}%
_{3}-2\sum_{j=1}^{3}Y^{k_{ji}}  \nonumber \\
N_{L_{i}} &=&(\frac{3}{2}r_{2}+\frac{3}{10}r_{1})\tilde{\alpha}%
_{3}-\sum_{j=1}^{3}Y^{l_{ij}}  \nonumber \\
N_{E_{i}^{c}} &=&(\frac{6}{5}r_{1})\tilde{\alpha}_{3}-2%
\sum_{j=1}^{3}Y^{l_{ji}}  \nonumber \\
N_{H_{U_{ij}}} &=&(\frac{3}{2}r_{2}+\frac{3}{10}r_{1})\tilde{\alpha}%
_{3}-3Y^{h_{ij}}  \nonumber \\
N_{H_{D_{ij}}} &=&(\frac{3}{2}r_{2}+\frac{3}{10}r_{1})\tilde{\alpha}%
_{3}-3Y^{k_{ij}}-Y^{l_{ij}}  \label{N}
\end{eqnarray}

The Yukawa RGEs are flavour independent and are driven to the flavour
independent infra-red stable fixed points (IRSFPs) of eq(\ref{YukRGEs}) 
\begin{eqnarray}
{R^{h}}^{*} &=&(\frac{232}{3}+45r_{2}+\frac{232}{15}r_{1}+15b_{3})/219 
\nonumber \\
{R^{k}}^{*} &=&(80+39r_{2}+\frac{21}{15}r_{1}+13b_{3})/219  \nonumber \\
{R^{l}}^{*} &=&(-24+54r_{2}+39r_{1}+18b_{3})/219  \label{FPs}
\end{eqnarray}
where ${R^{h}}^{*}\equiv \frac{{Y^{h}}^{*}}{\tilde{\alpha}_{3}}$, ${R^{k}}%
^{*}\equiv \frac{{Y^{k}}^{*}}{\tilde{\alpha}_{3}}$, ${R^{l}}^{*}\equiv \frac{%
{Y^{l}}^{*}}{\tilde{\alpha}_{3}}$, where ${Y^{h}}^{*}\equiv {Y^{h_{ij}}}^{*}$%
, ${Y^{k}}^{*}\equiv {Y^{k_{ij}}}^{*}$, ${Y^{l}}^{*}\equiv {Y^{l_{ij}}}^{*}$%
, $\forall i,j$. For example for $n=6$ we find $b_{3}=3$, $%
r_{1}=0.238,r_{2}=0.428$, ${R^{h}}^{*}=0.663$, ${R^{k}}^{*}=0.621$, ${R^{l}}%
^{*}=0.285$. (Note the approximate isospin symmetry in the IRSFPs.)

\section{Numerical Results}
In this section we examine the mass hierarchies and mixings which result 
from the democracy in the fixed points. We assume that the matrices 
$h_{ij}$,~$k_{ij}$ and $l_{ij}$ in Eq.(\ref{W}) correspond directly 
to the Yukawa couplings of the low energy MSSM below the scale $M_I$,
and that the fermion mass matrices are therefore given by 
\ba 
m_U &=& h_{ij} v_2\nn\\
m_D &=& k_{ij} v_1\nn\\
m_E &=& l_{ij} v_1
\ea
where $v_1$ and $v_2$ are the VEVs of the two light 
MSSM Higgs. 

We begin by making an ansatz for the 
mass matrices at the high $M_{NP}$ scale based on the 
$SU(3)_L\times SU(3)_R$ model -- \ie the matrices are 
parameterized by 5 parameters; 
\ba
\label{ansatz}
h_{ij} &=& (\delta_{i1}+a\delta_{i2}-(1+a)\delta_{i3})
(b_1\delta_{ji}+b_2\delta_{j2}-(b_1+b_2)\delta_{j3}) \nn\\
k_{ij} &=& c h_{ij} \nn\\
l_{ij} &=& d h_{ij}
\ea
(We shall be ignoring the question of CP violation here, although 
it could arise purely in the soft supersymmetry breaking terms 
as in Ref.\cite{me}.)
This form is the same as Eq.\ref{cups} in which the tracelessness 
arose because the Yukawa couplings were generated by the VEVs 
of adjoint fields.
This structure can be thought of as a kind of texture 
since the matrices have zero determinant at the $M_{NP}$ scale
and in fact have rank 1 -- \ie two zero eigenvalues apiece.
Generally this type of ansatz will always be required. The reason 
is that, although the RGEs in strong unified models can produce the 
third/second generation hierarchy, we still need to explain the 
smallness of the first generation. 

At this stage, one could be forgiven for thinking that we 
have not gained anything beyond what can be achieved with 
conventional texture models.
However in conventional texture models (based on the MSSM) 
the hierarchy we observe in the quark and lepton masses 
requires additional small parameters. This is because the RGEs of 
the MSSM dictate that the rank of the mass matrices is 
the same at all energy scales -- in this case, in the MSSM, there would 
always be six zero eigenvalues unless we invoked the Froggat--Nielsen 
mechanism. In strong unification however this is not the case because,
although the Yukawa couplings are drawn towards IRFPs which also 
have two zero eigenvalues, {\em these are not the same 
eigenvectors}. Hence there is a period 
during the running of the renormalisation group equations, 
before the Yukawa couplings have become significantly focused, 
in which the light quark masses receive small contributions. 
This is the origin of our small parameters and quite 
arbitrary values of the mass matrices at $M_{NP}$ result in 
masses and mixings in line with the observed pattern. 

We can see why (heuristically) by 
examining the form of the RGEs and approximating their 
solution by iteration. At the first step, \ie
\begin{eqnarray}
{Y^{h_{ij}}} &=&Y^{h_{ij}}(\mpl)
(1+\Delta t (N_{Q_{i}}+N_{U_{j}^{c}}+N_{H_{U_{ij}}})
)\nonumber \\
{Y^{k_{ij}}} &=&Y^{k_{ij}}(\mpl)
(1+\Delta t (N_{Q_{i}}+N_{D_{j}^{c}}+N_{H_{D_{ij}}})
)\nonumber \\
{Y^{l_{ij}}} &=&Y^{l_{ij}}(\mpl)(1+\Delta t (N_{L_{i}}+
N_{E_{j}^{c}}+N_{H_{D_{ij}}})),
\end{eqnarray}
we see that there is still one zero eigenvalue for each $Y$.
Hence the third/second 
and second/first generation mass-hierarchies are initially of order
${\Delta} t \sim few /16\pi^2 $.
Succesive iterations then tend to reduce the first and second 
generation masses reflecting the focussing effect of the fixed point. 
Further analytic proof that the Standard-Model-like structure is natural 
is difficult because the couplings are not immediately focussed. 

Henceforth we adopt an `empirical' approach: we begin 
at the unification scale with $\alpha_i=1$ but with the 5 remaining 
Yukawa parameters being chosen {\em randomly} around the central value,
\be
a=b_1=b_2=c=d=0
\ee
The parameters were varied randomly by a factor $g\times g_X$
(where $g_X=\sqrt{4\pi\alpha}$ is the gauge coupling at 
the unification scale) 
about these values, with $g$ going from $0\rightarrow 2$.
The running was stopped at $M_I$ ($\approx 10^8 \gev$ for 
$n=8$). In figures 1,2, and 3 we show the generated masses of the 
charm/up, strange/down and mu/electron 
for random values of the 5 parameters normalised to the 
running third generation masses at $m_t$.
(For proper comparison with the 
MSSM the parameters should of course be run from $M_I$ down to 
$m_t$ using the MSSM RGEs. Since the point here is simply to 
show that they are within the right range, this would 
not be relevant.)
In figure 4 we show the CKM parameters, $V_{us}$ and $V_{ub}$, and 
in figure 5 we show $V_{cb}$. 
(Since there is no CP violation in these Yukawa couplings, the 
corresponding CKM matrix depends on only these three parameters.)

The point of these diagrams is not to claim that we 
have a definite prediction for the fermion masses and mixings
(clearly we do not) but rather to show that the hierarchies observed 
arise naturally in the context of strong unification with a democratic Higgs.
In other words, without invoking any small parameters, Standard Model 
like hierarchies arise from rather ordinary choices 
of initial parameters. And indeed, all the results from our 
randomly selected initial parameters
are concentrated to within an order of magnitude about the 
correct physical values.
In this sense our approach is very different from the usual 
Froggatt-Nielsen picture, in which the hierarchies are `predicted' 
by the scale of breaking of some underlying symmetry.
It is also not difficult to find choices of 
parameters which give the physical values quoted 
by the Particle Data Group (modulo the MSSM running 
between $M_I$ and the weak scale). 

\section{Conclusion}

In this paper we have investigated the possibility that 
strong unification may be responsible for the generation
of fermion mass and mixing hierarchies. Given a democratic 
Higgs, the observed pattern appears to arise quite naturally. 
Some kind of ansatz is needed in order to obtain sufficiently light
first family masses, as in Eq.\ref{ansatz} which follows from
Eq.\ref{cups}.
After making this simple ansatz, which corresponds to zero determinant and 
two zero eigenvalues, our picture is very different from
the usual Froggatt-Nielsen `texture' approach in which further modification 
of the Yukawa matrices is controlled by powers of some expansion parameter.
The power of our approach was demonstrated by the fact that,
given this starting point, 
the remaining free parameters, {\em randomly chosen}, automatically give
rise to a spectrum of quark and lepton masses and mixing angles,
which has the character of the experimentally observed spectrum,
as seen in Figs.1-5. The success of this scheme is due to the
fact that the Yukawa matrices are strongly attracted towards
the infrared fixed point consisting of
democratic matrices with equal entries in every position.
Thus even though the matrices start out at high energies as
semi-random, they end up at low energies as quite accurately democratic.
Although this fixed point is quite rapidly approached, due to the
quickly falling gauge couplings of strong unification,
if we were to choose completely random matrices we would not arrive at
a spectrum resembling what is observed, and so something like the above ansatz
seems a necessary additional requirement.

Having achieved a set of approximately democratic Yukawa matrices at the
intermediate scale, we also demand that the two MSSM Higgs doublets
be extracted from the set of 18 high energy Higgs doublets 
in a democratic way. 
We presented two examples in which such 
a democratic Higgs may occur. The first has 
the virtue that it is based on the strong unification scenario
which already exists in the literature and has some motivation.
Unfortunately its complexity precludes a detailed analysis. 
Therefore we introduced a second model which gives some 
control over the generation of the democratic Higgs,
but in a perturbative framework which is clearly subject to possible
non-perturbative corrections beyond our control.
Finally we comment that other models, perhaps based on the 
more recent work of Ref.\cite{tony}, might be worth exploring.

\newpage

\begin{figure}
\vspace*{-1in}
\epsffile{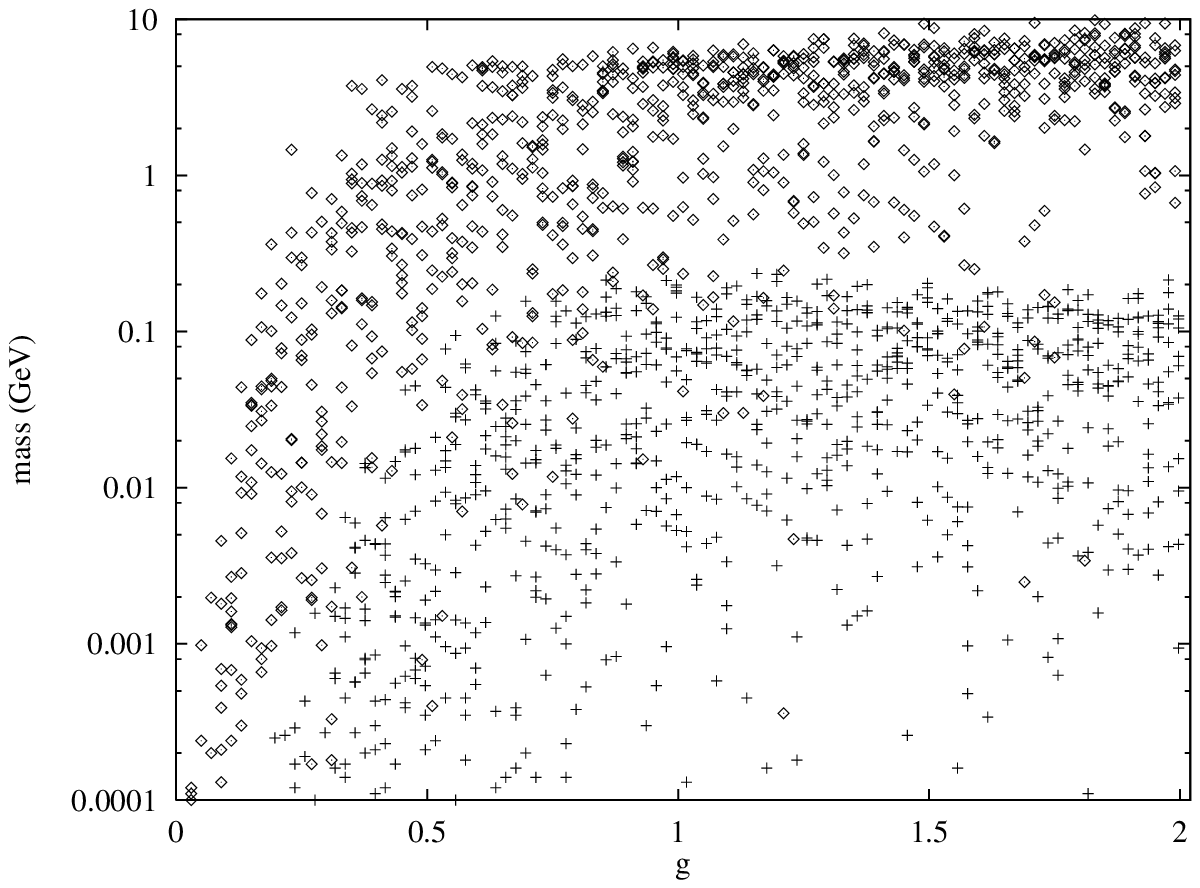}
\vspace{0.2in}
\caption{Charm and up masses at the intermediate scale 
normalised to $m_t(M_I)=160 \gev$.} 
\end{figure}

\begin{figure}
\vspace*{-1in}
\epsffile{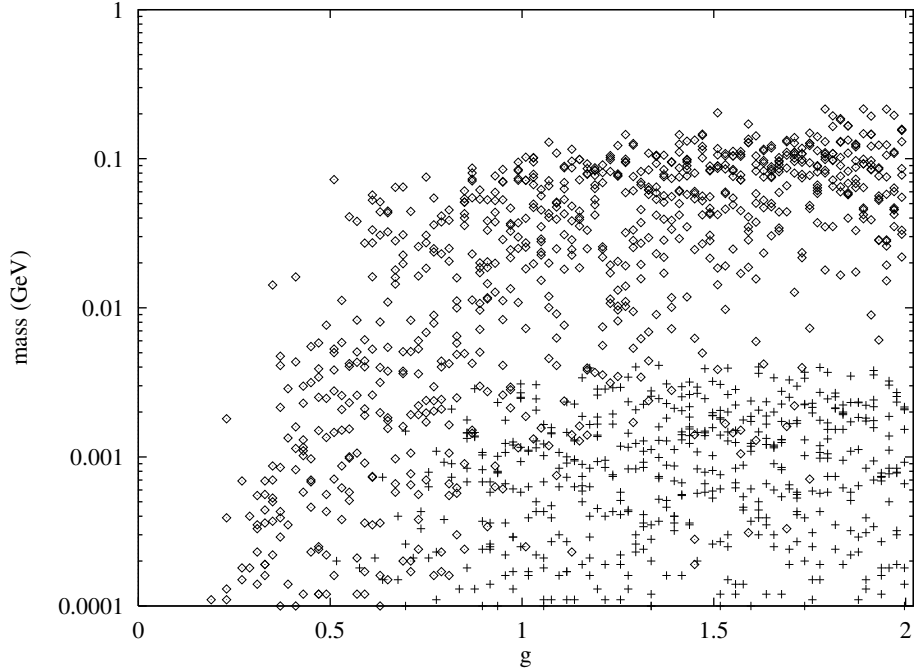}
\vspace{0.2in}
\caption{Strange and down masses at the intermediate scale, $M_I$, 
normalised to $m_b(M_I)=2.75\gev $.} 
\end{figure}

\newpage

\begin{figure}
\vspace*{-1in}
\epsffile{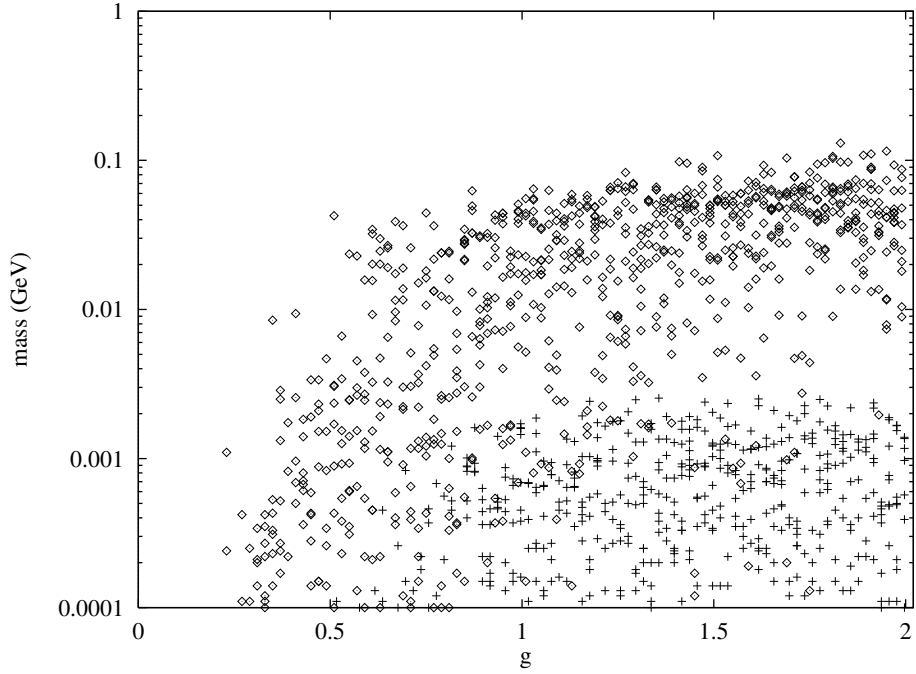}
\vspace{0.2in}
\caption{Mu and electron masses at the intermediate scale, $M_I$,
normalised to $m_\tau(M_I)=1.7\gev $.} 
\end{figure}

\begin{figure}
\vspace*{-1in}
\epsffile{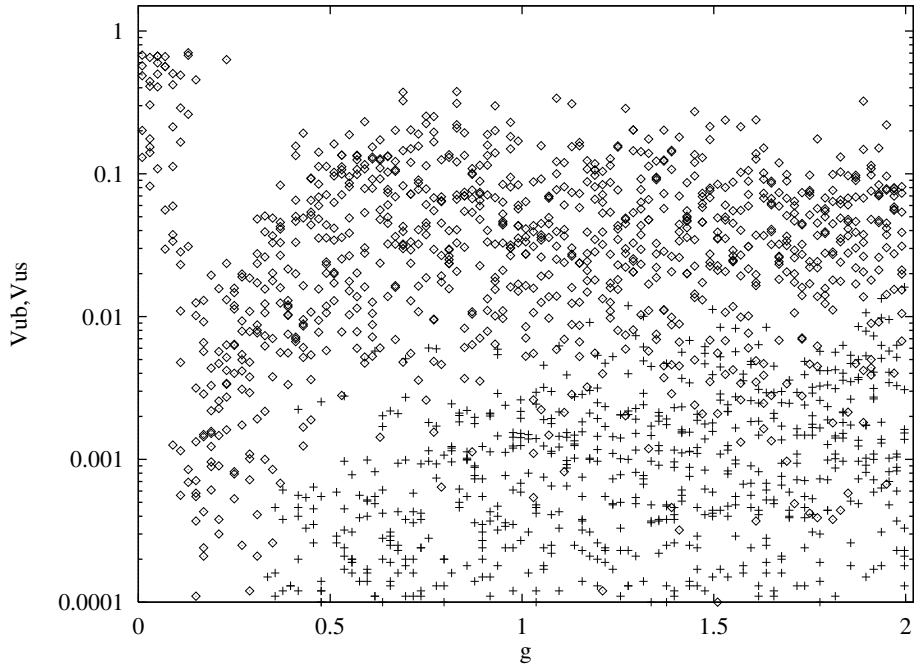}
\vspace{0.2in}
\caption{CKM phases $V_{us}$, $V_{ub}$ at the intermediate scale.} 
\end{figure}

\newpage

\begin{figure}
\vspace*{-1in}
\epsffile{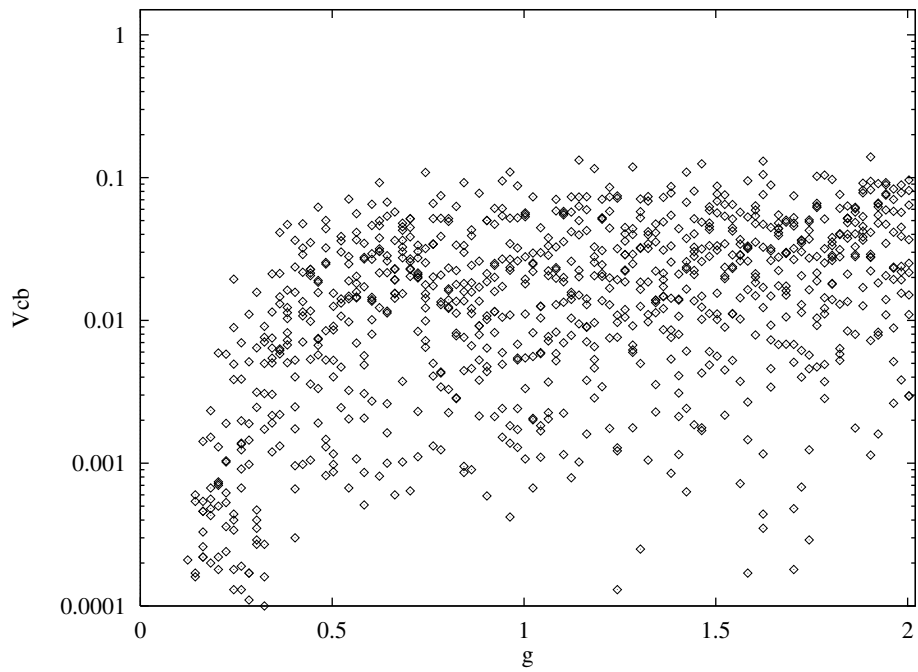}
\caption{CKM phase $V_{cb}$ at the intermediate scale.} 
\end{figure}

\end{document}